\documentclass[fleqn,usenatbib]{mnras}

\usepackage{newtxtext,newtxmath}

\usepackage[T1]{fontenc}

\DeclareRobustCommand{\VAN}[3]{#2}
\let\VANthebibliography\thebibliography
\def\thebibliography{\DeclareRobustCommand{\VAN}[3]{##3}\VANthebibliography}

\usepackage{graphicx}	
\usepackage{amsmath}	
\usepackage{booktabs}
\usepackage{siunitx}
\usepackage{multirow}

\title[]{The structure and evolution of the Galactic high-$\alpha$ disc I. Chemical and age orbital cartography}

\author[Akbaba et al.]{
Furkan Akbaba$^{1}$\thanks{E-mail: furkan.akbaba@ogr.iu.edu.tr},
Danny Horta$^{2}$,
Olcay Plevne$^{3}$
\\
$^{1}$Institute of Graduate Studies in Science, Istanbul University, Istanbul, Turkey\\
$^{2}$Institute for Astronomy, University of Edinburgh, Royal Observatory, Blackford Hill, Edinburgh, EH9 3HJ, UK\\
$^{3}$Faculty of Science, Department of Astronomy and Space Sciences, Istanbul University, Istanbul, Turkey
}

\date{Accepted XXX. Received YYY; in original form ZZZ}

\pubyear{\the\year{}}

\begin{document}
\label{firstpage}
\pagerange{\pageref{firstpage}--\pageref{lastpage}}
\maketitle

\begin{abstract}
We present a comprehensive chemical and age orbital cartography of the Galactic high-$\alpha$ disc using subgiant stars with precise ages, element abundances, and full phase-space information from the \textsl{LAMOST--Gaia} data set. Specifically, we map how average [Fe/H], [$\alpha$/Fe], and age vary across present-day kinematic and orbital coordinates. We analyse the data in full and across mono-abundance populations to measure element abundance-orbital and age-orbital gradients across orbital actions and angular-momenta. Our results show that the high-$\alpha$ disc exhibits clear and coherent gradients in [Fe/H], [$\alpha$/Fe], and age with orbits; these gradients are much stronger and sharper in orbital space than in present-day kinematics, showing that orbital diagnostics recover the intrinsic disc structure of old disc populations more effectively than instantaneous kinematic coordinates. We find that older high-$\alpha$ populations display qualitatively similar element abundance--orbital and age--orbital trends to stars in the low-$\alpha$ disc, although the high-$\alpha$ gradients are generally shallower. The presence of these ordered correlations indicates that the old high-$\alpha$ disc is structured, and preserved a strong fossil record of its early assembly despite the Milky Way's subsequent accretion history. This result implies that later mergers did not fully erase the chemical-orbital and age-orbital structure imprinted during the high-$\alpha$ disc's earliest formation epoch. All together, these findings indicate that the Galactic high-$\alpha$ disc formed mainly through inside-out and upside-down growth.
\end{abstract}





\section{Introduction}

\noindent Early studies of vertical star-counts in the Milky Way disc \citep{GilmoreReid1983} suggested that it is comprised by two kinematic components, a hot thick disc and a cold thin disc, and that these geometrically distinct populations are also distinguishable given their element abundance ratio or age distributions \citep{Gilmore1989,Wyse1995, Fuhrmann1998}.  With the advent of precise astrometric data from \textsl{Gaia} \citep{Gaia16} and large-scale spectroscopic surveys such as \textsl{APOGEE} \citep{Apogee}, \textsl{GALAH} \citep{Galah}, \textsl{LAMOST} \citep{LAMOST}, \textsl{DESI} \citep{Cooper2023}, it has become possible to study the Milky Way's stellar populations in unprecedented detail. While some work using these data corroborate the initial disc dichotomy hypothesis \citep{RecioBlancoetal2014,Wojno2016}, other works modelling larger samples of Milky Way disc stars in small bins of mono-abundance and mono-age populations (MAPs) have instead advocated for a smooth transition between the thick and thin components \citep{Bovyetal2012, Bovy2012b, Rix2013, Mackereth2017}, alleviating the need for two structurally independent populations. 

\begin{figure*}
    \centering
    \includegraphics[width=1\textwidth]{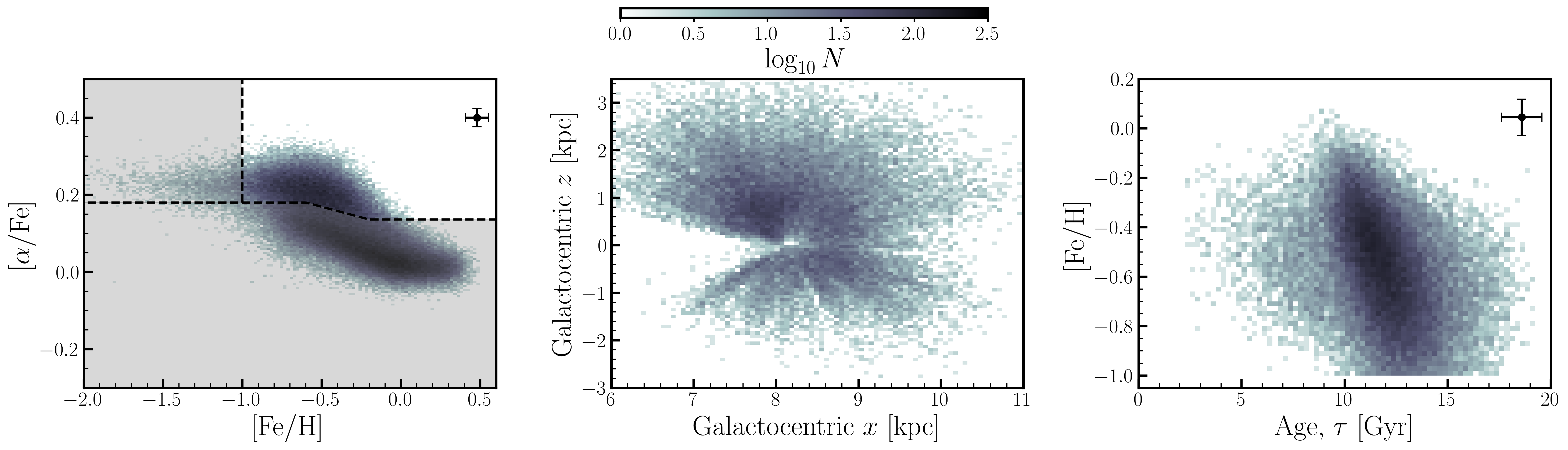}
    \caption{Chemical and spatial distributions of the stellar sample. The left panel shows the [$\alpha/$Fe]-[Fe/H] density map for the full dataset. The dashed curve indicates the adopted boundary used to define the high-$\alpha$ population. The grey region corresponds to [Fe/H] $\leq$ -1, which is not considered in the high-$\alpha$ selection. The middle and right panels show the spatial distribution ($x$--$z$) and the age-metallicity relation, respectively, for the stars classified as high-$\alpha$. Median uncertainties are shown in the corners of the panels. In the right panel, the ages of subgiant stars go above 14 Gyr due to the flat prior adopted in \citet{Xiang2022}. However, most of the high-$\alpha$ disc stars have $\tau<14$ Gyr, and since this work focuses on modelling the median value across orbital properties, this tail of older stars do not affect our results. }
    \label{fig:data_intro}
\end{figure*}

Reconciling these two lines of thought requires understanding the peculiar element abundance ratios of Milky Way disc stars in the [$\alpha$/Fe]-[Fe/H] plane. It is widely accepted that the Galactic disc displays two chemically distinct sequences: a high-$\alpha$ sequence and low-$\alpha$ sequence \citep[see the left panel of Figure~\ref{fig:data_intro} and][]{Bensby2003, Adibekyan2012, Bensby14, Nidever2014,Hayden2015,Plevne2020}; for stars around the Sun, these two sequences have been shown to overlap in [Fe/H], leading to an ``$\alpha$-bimodality''. The two chemical disc sequences are referred to as the high-$\alpha$ and low-$\alpha$ discs, respectively, and are commonly linked to the geometrically defined thick and thin discs based on the properties of their constituent stars. This is because stars in the high-$\alpha$ disc are predominantly older and kinematically hotter when compared to those in the low-$\alpha$ one \citep{Mackereth2019,Queiroz2020, Lian2022,Imig2023}, analogous to the kinematic thick/thin disc separation proposed in early studies.

Characterising the formation of the high-$\alpha$ disc is crucial to understand the Galaxy's early stages of formation and the genesis of its disc. Observational studies have shown that this component is comprised of stars that are $\alpha$-enhanced ([$\alpha$/Fe]$\gtrsim0.15$ dex), have sub-solar metallicities ($-1\lesssim$[Fe/H]$\lesssim0$ dex), and are kinematically hotter than their low-$\alpha$ counterparts ($v_{\phi}\lesssim150$ km s$^{-1}$). The properties of stars pertaining to the high-$\alpha$ disc provide signatures of rapid star formation on short timescales in a turbulent, gas-rich Galaxy at early cosmic times \citep[][]{Xiang2022, Belokurov2022,Chandra2024,Zhang2024,Viswanathan2025}. However, the precise formation mechanism needed to explain the observational data for this population remains unresolved. Competing scenarios include (i) vertical heating of a pre-existing thin disc through minor mergers \citep{Villalobos2008, Bird2013}, (ii) in-situ formation during a turbulent, high-redshift phase \citep{Brook2004, Bournaud2009}, and (iii) radial migration and secular processes within the disc \citep{Schoenrich2009, Minchev2010}. Recent simulations have also suggested that high-$\alpha$ sequences may arise naturally in gas accretion and bursty star formation environments \citep{Grand2018, Agertz2021, Renaud2025}.

In the first of a series of articles, we set out to constrain the formation scenario of the Milky Way's high-$\alpha$ disc by examining the trends of element abundance ratios and age with respect to kinematic and dynamical properties using data delivered by large-scale surveys. In this paper, we perform a comprehensive chemical and age orbital cartography synergising spectroscopic data for subgiant stars from the \textsl{LAMOST} survey with \textsl{Gaia DR3} astrometry and ages determined by \citet{Xiang2019}. In detail, we investigate the properties of these stars in several orbital planes to trace how key elemental abundance ratios (e.g., [Fe/H] or [$\alpha$/Fe]) and age vary across different dynamical diagrams. The gradients derived in these planes provide insights into how chemistry and stellar birth dates (i.e., approximately time-invariant quantities) correlate with orbital structure, offering new constraints on the structure and evolution of the high-$\alpha$ disc.

This paper is organized as follows: Section~\ref{sec:data} describes the data and sample selection. Section~\ref{sec:results} presents the results of our chemical-orbital analyses. Section~\ref{sec:discussion} discusses the implications of our findings for the formation and evolution of the high-$\alpha$ population. Finally, Section~\ref{sec:conclusions} summarises our conclusions.

\section{Data}
\label{sec:data}

We use the public catalog of subgiant stars computed by \citet[][]{Xiang2022} using the \textsl{LAMOST} and \textsl{Gaia} surveys. This catalog comprises stellar parameters and element abundances computed using the data-driven Payne \citep{Xiang2019}, stellar ages computed using colour-magnitude diagram fitting, and the required 6D phase-space information to determine kinematics/orbits from \textsl{Gaia}/\textsl{LAMOST} observables. In total, this sample contains $\approx250{,}000$ subgiant stars. 

We determine kinematics and orbits for these stars using the line-of-sight velocities from \textsl{LAMOST} data together with the on-sky positions, proper motions, and distances (estimated as the inverse of the parallax) from \textsl{Gaia}. We transform these observed quantities to Galactocentric Cartesian coordinates using the \texttt{Astropy} library \citep{Astropy2013, Astropy2018}, adopting a solar position in the Milky Way disc of $R_{\odot} = 8.275$ kpc \citep{Gravity2021} and $z_{\odot} = 0.02$ kpc \citep{Bennett2019}, and assuming a local standard of rest velocity of $V_{\mathrm{LSR}} = 232.8$ km s$^{-1}$ and the Solar peculiar motion with respect to the LSR as $(U, V, W) = (11.10, 12.24, 7.25)$ km s$^{-1}$ \citep{Schonrich2010}. We then also integrate stellar orbits using the \texttt{galpy} package \citep{galpy} assuming the \texttt{Cautun2020} Galactic potential \citep{Cautun2020}; we compute orbital actions using the St\"ackel Fudge approximation, and empirically calculate a star's (Cartesian) angular momentum components ($L_x,L_y, L_z$) via $\vec{L} = \vec{v} \times x$.

Lastly, we employ the isochrone-based ages from \citet[][]{Xiang2022} that are computed using the Yonsei-Yale isochrones (see \citet[][]{Xiang2022} for more details). All together, this leads to a sample of (subgiant) stars with reliable element abundance, kinematic/orbit, and age information.

Our parent working sample is comprised of stars that satisfy the following selection criteria:

\begin{enumerate}
    \item[\textbf{Medium signal-to-noise spectra:}] \textsl{LAMOST} spectral S/N $> 20$,
    \item[\textbf{Subgiant stars:}] \textsl{LAMOST}-determined atmospheric parameters, effective temperature and surface gravity between $4800<T_{\mathrm{eff}}<6000$ K and $2 < \log(g) < 5$ dex,
    \item[\textbf{Nearby high-$\alpha$ disc stars:}] $d<4$ kpc and stars that fall in the high-$\alpha$ sequence of the [$\alpha$/Fe]-[Fe/H] plane (see Fig. \ref{fig:data_intro}). These stars are defined as those satisfying
    $\mathrm{[Fe/H]} \geq -1$ and lying above the separation line defined as follows:
    
\[
[\alpha/\mathrm{Fe}] =
\begin{cases}
0.18, & [\mathrm{Fe/H}] < -0.6, \\
-0.11\times [\mathrm{Fe/H}] + 0.114, & -0.6 \leq [\mathrm{Fe/H}] \leq -0.2, \\
0.136, & [\mathrm{Fe/H}] > -0.2 .
\end{cases}
\] 
    \item[\textbf{Precise age measurements:}] $\mathrm{age}/\sigma_{\mathrm{age}}\geq5$, 
    \item[\textbf{High-quality astrometry and distances:}]   \textsl{RUWE} $<1.2$ and $\sigma_d/d<0.2$,

\end{enumerate}

These quality cuts yield a sample of 45,335 high-$\alpha$ stars shown in Fig~\ref{fig:data_intro}. We choose to limit our sample to stars within $d<4$ kpc to ensure that the parallax measurements yield good distance estimates.

\section{Chemical and age orbital cartography of the high-$\alpha$ disk}
\label{sec:results}

\begin{figure*}
    \centering
    \includegraphics[width=1\textwidth]{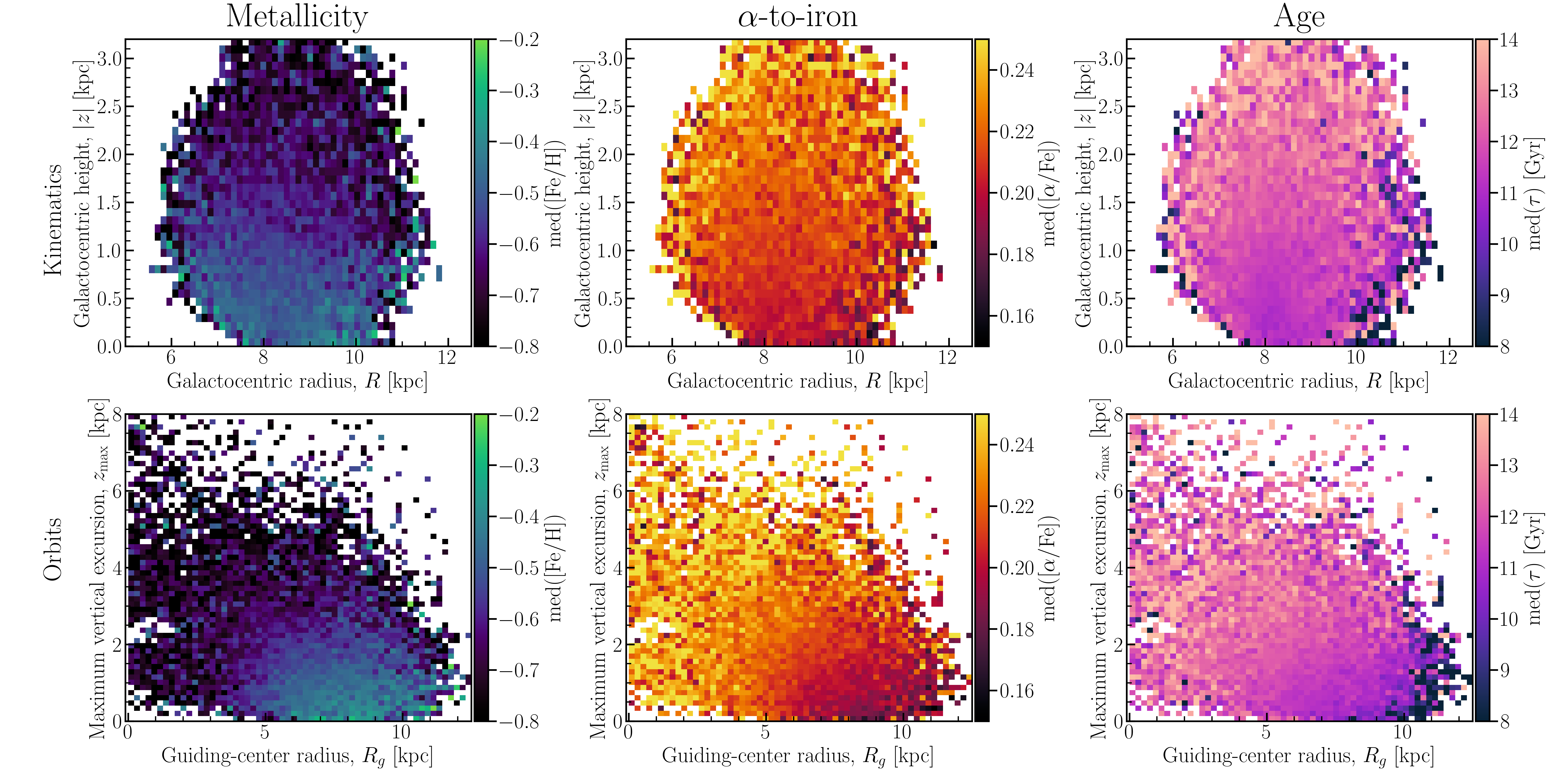}
    \caption{Kinematic and orbital distributions of the high-$\alpha$ population. The top row shows maps in the $R$-$|z|$ plane, while the bottom row displays the corresponding distributions in the $R_g$-$z_{\max}$ plane. From left to right, panels are colour-coded by median of [Fe/H], [$\alpha$/Fe], and stellar age. The high-$\alpha$ disc shows clear element abundance and age trends as a function of kinematics and orbits, which resemble that of stars in the low-$\alpha$ disc (see Appendix~\ref{app_lowalpha}). However, these trends are much stronger as a function of orbital properties.}
    \label{fig:kinematicsvsorbits}
\end{figure*}

Interpreting element abundance and age trends across the Galactic disc requires examining both present-day phase-space coordinates and quantities that better preserve information about a star's orbit. Kinematic quantities provide an instantaneous measurement of a star's position and motion, but do not retain a time-integrated record of the motion of such star around the Galaxy. Conversely, the dynamical properties of a star, manifested via its orbital parameters, provide a time-averaged description of its stellar orbit, and thus contain information about the mass distribution. Motivated by this distinction, we compare element abundance and age distributions in the high-$\alpha$ disc using a set of kinematic and orbital diagrams.

Figure~\ref{fig:kinematicsvsorbits} illustrates the comparison between trends of element abundance and age w.r.t. kinematics or orbits for the high-$\alpha$ disc population. The top panels display maps using present-day Galactocentric radius, $R$, and absolute Galactocentric height, $|z|$, while the bottom panels show the same stars as a function of their guiding-center radius, $R_g$, and maximum orbital vertical excursion, $z_{\mathrm{max}}$. The guiding-center radius values were computed assuming a flat rotation curve using $R_g \approx L_z/v_c$, where $L_z$ is the azimuthal component of the angular momentum and $v_c$ is the circular velocity curve of the Milky Way, taken from \citet{Eilers2019}. While $R_g$ is primarily employed for stars on near-circular orbits, we use it here as a test to compare orbits with kinematics, and it is used for qualitative comparison purposes only. Each panel is pixelated into 60 pixels per side, where each pixel illustrates either med([Fe/H]) (left), med([$\alpha$/Fe]) (centre), or median value of stellar age (med($\tau$), right) for all stars that fall in that pixel. We choose to only show average [Fe/H] and average [$\alpha$/Fe] as these trace the contribution from core-collapse (SN II) and white-dwarf (SN Ia) supernovae, two of the main drivers of star formation in galaxies; moreover, these are two of the more well measured element abundance ratios in the subgiant catalogue.

Comparing the top and bottom rows of Figure~\ref{fig:kinematicsvsorbits}, we see that there are some immediate differences with respect to the extent of the data in kinematic or orbit space. For example, in the top row ($R$--$|z|$ plane), we find that stars are concentrated within $7 \lesssim R\lesssim 10$ kpc and $|z|<3$ kpc. Conversely, the orbital excursions of stars cover a broader region of the Galaxy, spanning from $0 \lesssim R_g \lesssim 11$ kpc and $z_{\mathrm{max}}<6$ kpc. 

When examining the gradient of median element abundance or median age in Figure~\ref{fig:kinematicsvsorbits} we find three important results: 1) there is a trend in median element abundance and median age, and these trends appear qualitatively the same; 2) the trend of med([Fe/H]) goes in the opposite direction than the trend in med([$\alpha$/Fe]); 3) the trends of median element abundance and median age appear stronger in orbits than in kinematics. 

In more detail, when inspecting the plane of kinematics, med([Fe/H]) and med([$\alpha$/Fe]) exhibit a weak negative(positive) gradient with $|z|$, which is approximately equal across all $R$; neither of these abundances show a radial dependence. This vertical dependence and radial independence is also seen in mean age. On the contrary, in orbit space, med([Fe/H]) shows a clear negative gradient with $z_{\mathrm{max}}$ and a positive gradient with $R_{g}$, implying that stars on more vertically extended orbits and with smaller guiding-centre radii are systematically more metal-poor. Interestingly, the gradient in med([$\alpha$/Fe]) is flipped when compared to med([Fe/H]), but the structure overall remains the same, leading to a positive vertical and negative radial gradient w.r.t. orbits. When inspecting the median age distribution, we find that stars on more vertically extended orbits, which on average are more [Fe/H]-poor and [$\alpha$/Fe]-enhanced, are also older. 

We have checked to ensure that the gradients we see aren't a result of internal scatter (Fig~\ref{fig:kinematics_vs_orbits_MAD}). We find that the average scatter value is approximately the same across all pixels for the MAD of [Fe/H], [$\alpha$/Fe], and age. These values are relatively small, on the order of $\approx0.13$ dex for [Fe/H], $\approx0.02$ dex for [$\alpha$/Fe], and $\approx1$ Gyr for age.

In summary, while kinematic maps reveal only shallow element abundance and age gradients, orbital maps uncover clear gradients in the high-$\alpha$ disc. These results highlight the advantage of orbital-based analyses for isolating the intrinsic chemical-orbital structure of older, more phase-mixed, and more dynamically heated populations like the high-$\alpha$ disc. 

\begin{figure*}
    \centering
    \includegraphics[width=1\textwidth]{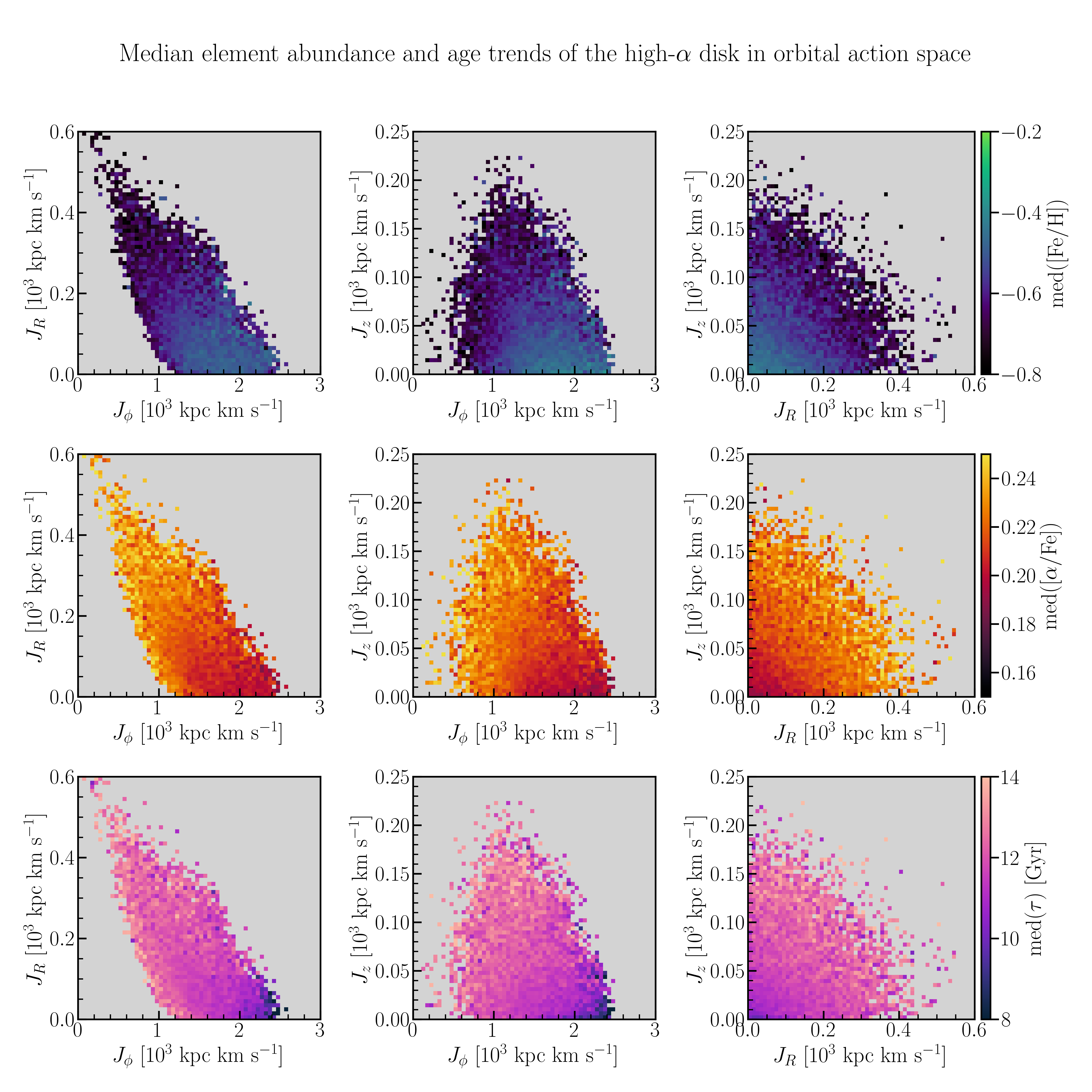}
    \caption{Distribution of high-$\alpha$ stars in action space. The panels show the median values of [Fe/H] (top row), [$\alpha$/Fe] (middle row), and stellar age (bottom row) across different projections of the actions $(J_R, J_z, J_\phi)$. Colors indicate the mean chemical abundance or age within each bin, highlighting the element abundance--orbital and age--orbital structure of the high-$\alpha$ disc population.}
    \label{fig:subgiant_actions}
\end{figure*}
\subsection{Element abundance and age gradients as a function of orbits}
\subsubsection{Actions}
\label{sec_actions}

Following our analysis from the previous section, where we have learned that element-abundance and age gradients in the high-$\alpha$ disc appear clearer in orbit space than in kinematic diagrams, in this Section we go on to examine how median element abundance and median age trends in the high-$\alpha$ disc appear in orbital actions and angular momentum.

\begin{figure*}
    \centering
    \includegraphics[width=1\textwidth]{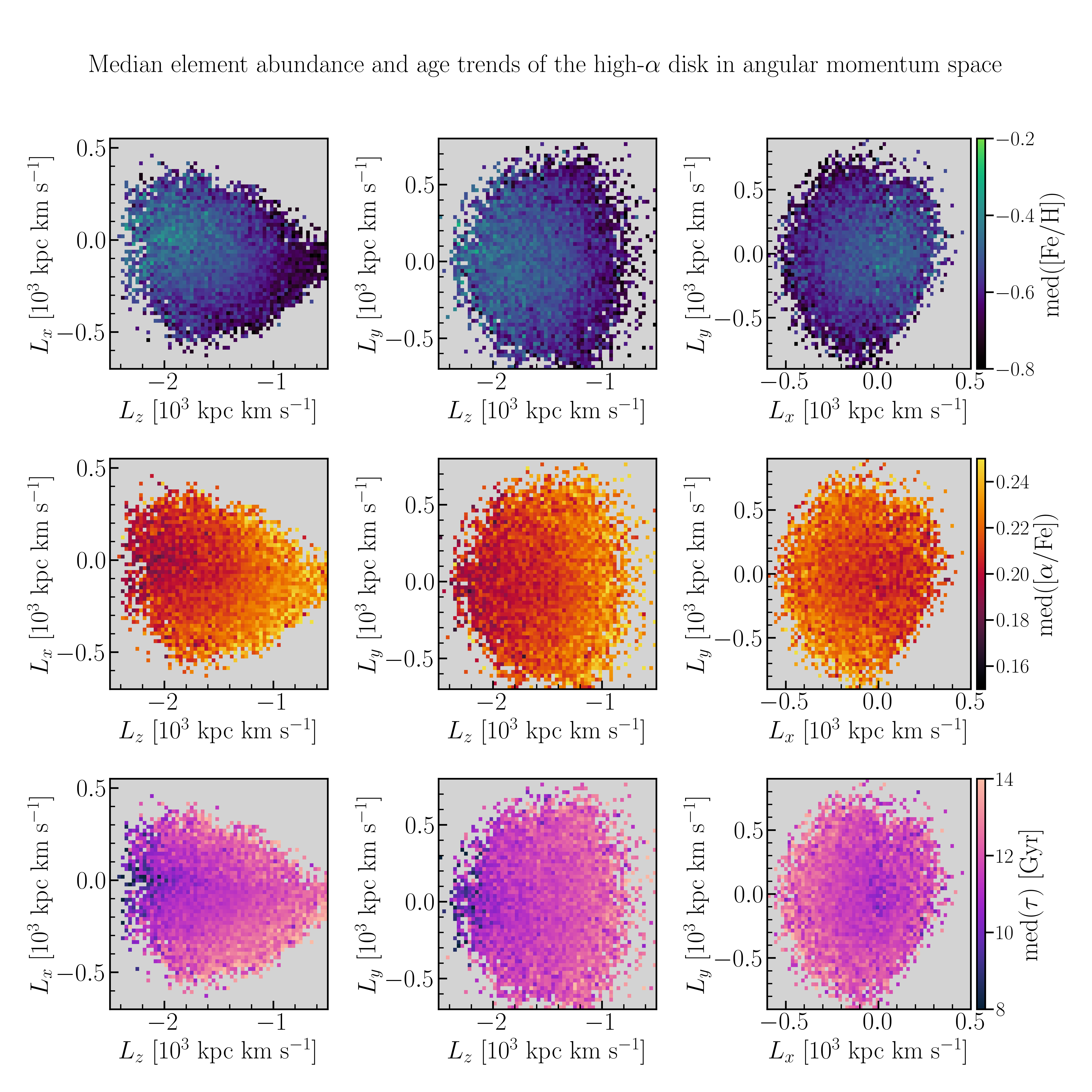}
    \caption{Distribution of high-$\alpha$ stars in angular momentum space. The panels show the median values of [Fe/H] (top row), [$\alpha,$/Fe] (middle row), and stellar age (bottom row) across different projections of the angular momentum components $(L_x, L_y, L_z)$. Colors indicate the median chemical abundance or median age within each bin, highlighting the element abundance--orbital and age--orbital structure of the high-$\alpha$ disc population.}
    \label{fig:subgiant_angularmomentum}
\end{figure*}

Orbital action variables offer a more general and physically motivated framework for characterising stellar dynamics \citep{Binney2008}. In a static axisymmetric system, actions are adiabatic invariants and are therefore expected to preserve information about stellar orbits. To compute these, we use the St\"akel approximation in \texttt{galpy} \citep{galpy}; for the vast majority of disc orbits, this approximation provides a more cost-efficient way of computing accurate orbital actions \citep{Sanders2016}\footnote{We also conducted an experiment to assess how well the assumption of axisymmetry and steady-state holds for old high-$\alpha$ disc stars, to ensure it is viable to use orbital actions. To do so, we computed the components of the angular momentum in two ways: empirically (see Section~\ref{sec:data}), and by integrating the orbits of stars. We then assessed how well these match, finding a remarkable level of agreement. Thus, we are confident that orbital actions will describe the orbits of high-$\alpha$ disc stars well.}.

Figure~\ref{fig:subgiant_actions} presents the distribution of high-$\alpha$ stars in orthogonal projections of radial, azimuthal, and vertical actions (namely, $J_R, J_\phi$, and $J_z$). The left column shows the variation in radial and azimuthal actions, the middle column shows the variation in vertical and azimuthal actions, and the right column shows the variation across radial and vertical actions. As shown in Figure~\ref{fig:kinematicsvsorbits}, we pixelate each plane and show the median values of [Fe/H] (top row), [$\alpha$/Fe] (middle row), and stellar age ($\tau$, bottom row). 

Focusing on the top row of Figure~\ref{fig:subgiant_actions}, we see that the med([Fe/H]) distribution shows clear and coherent gradients across action space. med([Fe/H]) exhibits a negative gradient with increasing $J_R$ or $J_z$, such that stars on less circular orbits (i.e., have larger radial and vertical excursions) are systematically more [Fe/H]-poor. In contrast, med([Fe/H]) displays a positive gradient with $J_\phi$, with higher med([Fe/H]) associated with larger angular momentum, $J_\phi$ (or $L_z$); this is what is also seen in the low-$\alpha$ disc (see Fig \ref{fig:lowalpha_action}), albeit with a stronger gradient \citep{Minchevetal2014, Frankel2018}. These trends are smooth and continuous, closely mirroring those seen in the orbital $R_{g}$--$z_{\mathrm{max}}$ plane, and indicate a strong coupling between med([Fe/H]) and orbital heating in the high-$\alpha$ disk. 

If one now focuses on the middle row, it becomes apparent that med([$\alpha$/Fe]) displays a somewhat similar behavior to med([Fe/H]). However, the gradient with med([$\alpha$/Fe]) flips direction when contrasted to med([Fe/H]), as also seen in the $R_g$--$z_{\mathrm{max}}$ plane (Figure~\ref{fig:kinematicsvsorbits}). Here, med([$\alpha$/Fe]) increases with increasing $J_R$ or $J_z$. Along those lines, med([$\alpha$/Fe]) decreases with increasing $J_\phi$, yielding a negative azimuthal gradient. Despite the opposite directions of the gradient between med([Fe/H]) and med([$\alpha$/Fe]) with orbital actions, the distribution of the data appears indistinguishable; in other words, regions of action space with unique med([Fe/H]) also have unique med([$\alpha$/Fe]).

Lastly, in the bottom row of Figure~\ref{fig:subgiant_actions} we examine the median age gradient in action space. We find that there is a gradient in age with orbital actions. Here, stars on orbits with larger radial and vertical actions are typically older, whilst stars on more circular orbits (and thus low $J_R$ and $J_z$ but high $J_\phi$) are younger. The age range probed spans $\approx6$ Gyr and includes some of the oldest stars in the Milky Way, reaching an average age of $\tau\sim14$ Gyr. Furthermore, the trend in average age follows the same profile as the one seen in med([Fe/H]) and med([$\alpha$/Fe]). The scatter in the values of average element abundance ratios and age for each mono-orbit population is low (Fig~\ref{fig:subgiants_actions_MAD}); this finding indicates that the trends in average element abundance ratios and age as a function of orbits attest for a high-$\alpha$ disc population that is structured.

\subsubsection{Angular momentum}

Following the results obtained in action space, in this Section we examine how the element abundances and stellar ages of high-$\alpha$ disc stars vary as a function of an orbital quantity, that unlike actions, does not require a parameterisation of the Milky Way potential: angular momentum. We compute the components of the angular momentum vector for each high-$\alpha$ disc star ($L_x$, $L_y$, $L_z$) using $\vec{L} = x \times \vec{v}$.

Figure~\ref{fig:subgiant_angularmomentum} shows the angular momentum equivalent of Figure~\ref{fig:subgiant_actions}. Here, the left column shows (Galactocentric) $L_x$--$L_z$, the middle column shows $L_y$--$L_z$, and the right column shows $L_x$--$L_y$; the top row illustrates the high-$\alpha$ disc sample pixelated by median [Fe/H], in the middle row by median [$\alpha$/Fe], and in the bottom row by median stellar age. 

The top and middle rows of Figure~\ref{fig:subgiant_angularmomentum} reveal that the element abundance gradients observed in action space (Figure~\ref{fig:subgiant_actions}) are also present in angular momentum space. Similarly to when examining the trends with respect to orbital actions, we find that med([Fe/H]) and $\langle \tau \rangle$ show a positive gradient with increasing $L_z$ (from left to right), while med([$\alpha$/Fe]) shows a negative gradient (from right to left). Interestingly, the $L_x$--$L_y$ plane shows that more med([Fe/H])-rich, med([$\alpha$/Fe])-poor, and younger high-$\alpha$ disc populations concentrate at lower $L_x$ and $L_y$, highlighting that on average more younger and chemically enriched populations tend to have less heated orbits; the morphology of mono-abundance contours found in this plane is similar to the ones seen in $z$--$v_z$ space for the low-$\alpha$ disc \citep{Price2021, Horta2024, Horta2026}. The scatter in these element abundance ratios and age is small per pixel (Fig~\ref{fig:subgiants_angular_momentums_MAD}).

In summary, Figure~\ref{fig:subgiant_angularmomentum} demonstrates that element abundance gradients in the high-$\alpha$ disc also appear in angular momentum space. These results indicate that angular momentum-based analyses support the action-based findings, and provide a complementary avenue for empirically examining element abundance or age orbital gradients that do not require the parameterisation of the Galactic potential or orbit integration.

\subsection{Chemical and age orbital cartography with MAPs}
\label{sec_maps}

In the previous Sections, we have shown that element abundance and age gradients in the high-$\alpha$ disc become significantly clearer when examined in orbital space rather than in kinematic diagrams. Here, we extend this analysis by adopting a mono-abundance population (MAP) framework \citep[e.g.,][]{Bovy2012b}, which allows us to isolate small groupings of stars sharing a common chemical profile (and given our results from the previous section, by default a similar age), thus likely sharing a common (more localised) formation environment. 

In practice, the MAPs we use are defined as narrow bins in the [Fe/H]-[$\alpha$/Fe] plane, with a bin size of $0.05 \times 0.025~\mathrm{dex}^2$ (the average uncertainty in [Fe/H] is $0.05$ dex, whilst for [$\alpha$/Fe] is $0.02$ dex). By construction, this approach provides a physically motivated way of tracing the underlying chemical-age-orbital structure of the high-$\alpha$ disc using stars that have formed in the same environment or under the same chemical enrichment pathway.

\subsubsection{Actions}

\begin{figure*}
    \centering
    \includegraphics[width=1\textwidth]{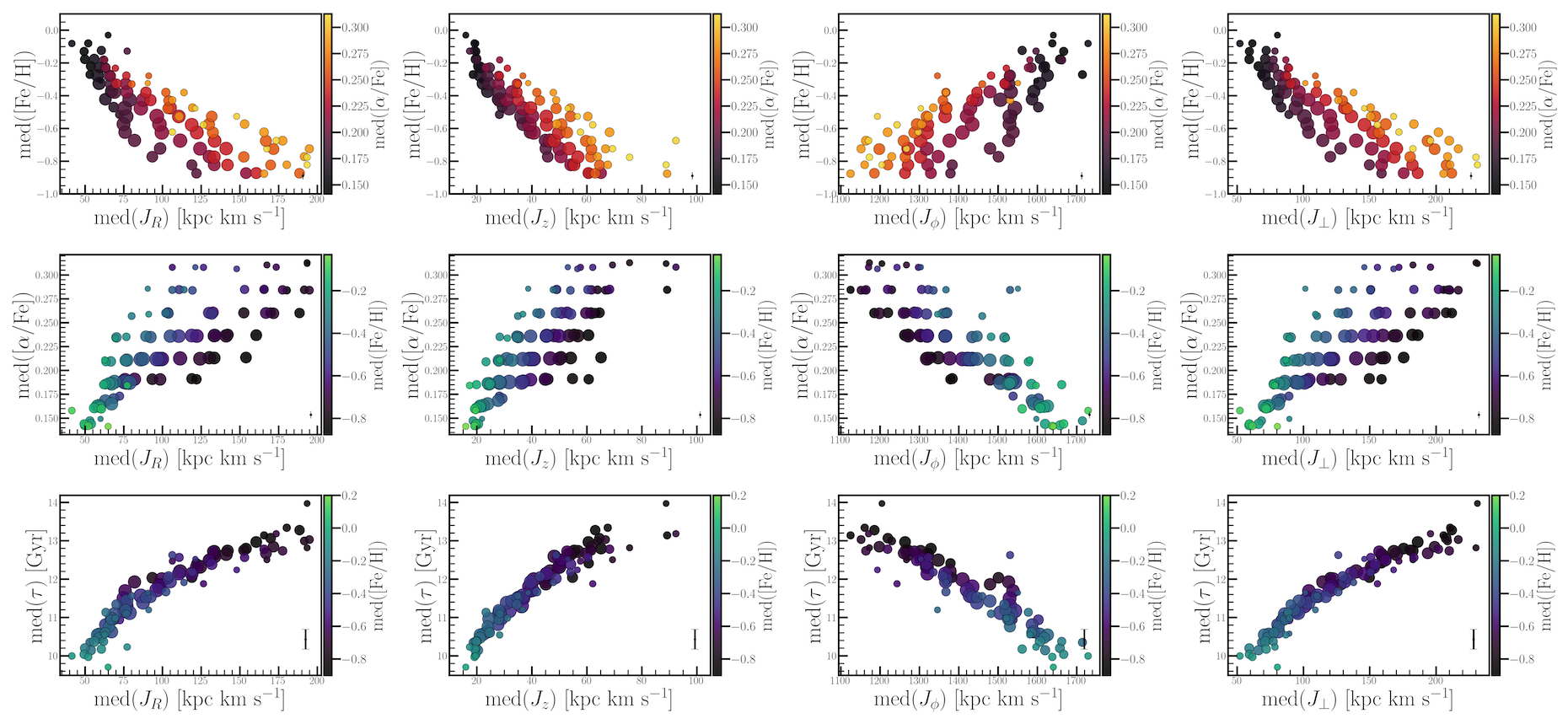}
    \caption{Correlations between element abundances, stellar ages, and actions for high-$\alpha$ disc stars. The top row shows the relation between [Fe/H] and the actions $J_R$, $J_z$, $J_\phi$, and $J_\perp$, while the middle row presents the corresponding relations for [$\alpha$/Fe]. The bottom row displays the dependence of stellar age $\tau$ on the same actions. Marker colors indicate the complementary abundance ([$\alpha$/Fe] in the top row, [Fe/H] in the middle row) and the [Fe/H] ratio in the bottom row, while marker sizes scale with the number of stars in each mono-abundance bin. Typical uncertainties of the median values, estimated as $\sigma/\sqrt{n}$ for each mono-abundance bin, are shown as representative error bars in the lower-right corner of each panel. See the text in Section\ref{sec_maps} for details.}
    \label{fig:subgiants_actions_map}
\end{figure*}

Figure~\ref{fig:subgiants_actions_map} presents the MAPs results for element abundances and stellar ages as a function of orbital actions for high-$\alpha$ disc stars. Each point in every panel shows the median value, where larger points correspond to MAPs that contain more stars. In detail, Figure~\ref{fig:subgiants_actions_map} illustrates how average [Fe/H] (top row), average [$\alpha$/Fe] (middle row), and average stellar age (bottom row) vary as a function of the radial action (left column), the azimuthal action (center-left column), the vertical action (center-right column), and the perpendicular action\footnote{The perpendicular action defined as $J_\perp = \sqrt{J_R^2 + J_z^2}$.} (right column). Marker colors represent the complementary element abundance ([$\alpha$/Fe] in the top row, [Fe/H] in the middle row, and [Fe/H] in the bottom row). We choose to only show MAP bins containing 30 stars or more.

The top row of Figure~\ref{fig:subgiants_actions_map} shows that the median value of [Fe/H] decreases approximately linearly with increasing $J_R$ and $J_z$. As expected, this leads to a negative trend between med([Fe/H]) and $J_\perp$. This result indicates that more [Fe/H]-poor ensembles of stars are on more heated orbits (i.e., more extended radial and vertical orbits), on average. Conversely, med([Fe/H]) shows a linearly positive correlation with $J_\phi$, albeit some scatter; stars with higher angular momentum tend to be more [Fe/H]-rich, on average.

The middle row of Figure~\ref{fig:subgiants_actions_map} shows that the median [$\alpha$/Fe]--action relations are inverted with respect to those found for [Fe/H]: average [$\alpha$/Fe] increases with $J_R$ and $J_z$ (and consequently with $J_\perp$), yet decreases with increasing $J_\phi$. These inverted gradients reflect the differing roles of core-collapse (SN II) and Type Ia supernovae in shaping the chemical enrichment of the high-$\alpha$ disc, and their connection to orbital structure. Notably, the [$\alpha$/Fe] distribution does not mirror [Fe/H] in a one-to-one fashion; instead, it displays a ridge-like structure in action space, with a more pronounced flattening at intermediate action values. This behavior is consistent with the fact that [$\alpha$/Fe] primarily traces star-formation timescales and saturates once the SN Ia contribution becomes dominant, whereas [Fe/H] reflects cumulative enrichment and varies more continuously with the degree of orbital heating. 

The bottom row shows the relation between median stellar age and orbital actions. Overall, we find that the trends of orbital action with age for MAPs appears to be the strongest. Interestingly, the trends with median age follow a similar profile to those seen in med([$\alpha$/Fe]), but however with much lower scatter. Here, younger groupings of stars tend to have lower radial, vertical, and perpendicular action magnitudes when compared to their older counterparts; similarly, on average younger stars tend to have higher $J_\phi$, and are thus on more circular orbits when compared to older stars.

The strong gradients in med([Fe/H]), med([$\alpha$/Fe]), and med(age) with orbital actions seen in MAPs for high-$\alpha$ disc stars suggests that this population is structured, and reinforces the the picture that the chemical enrichment of the high-$\alpha$ disc is closely linked to the degree of orbital heating. We have quantified these trends applying a weighted linear regression to the MAP-averaged median element abundance and age values, where the weights are determined by the number of stars in each MAP bin and the corresponding abundance uncertainties. The resulting slope and bias values are listed in Table~\ref{tab:gradients}, where the uncertainties on the linear model parameters are computed using ordinary least-squares regression.

\subsubsection{Angular momentum}

\begin{figure*}
    \centering
    \includegraphics[width=1\textwidth]{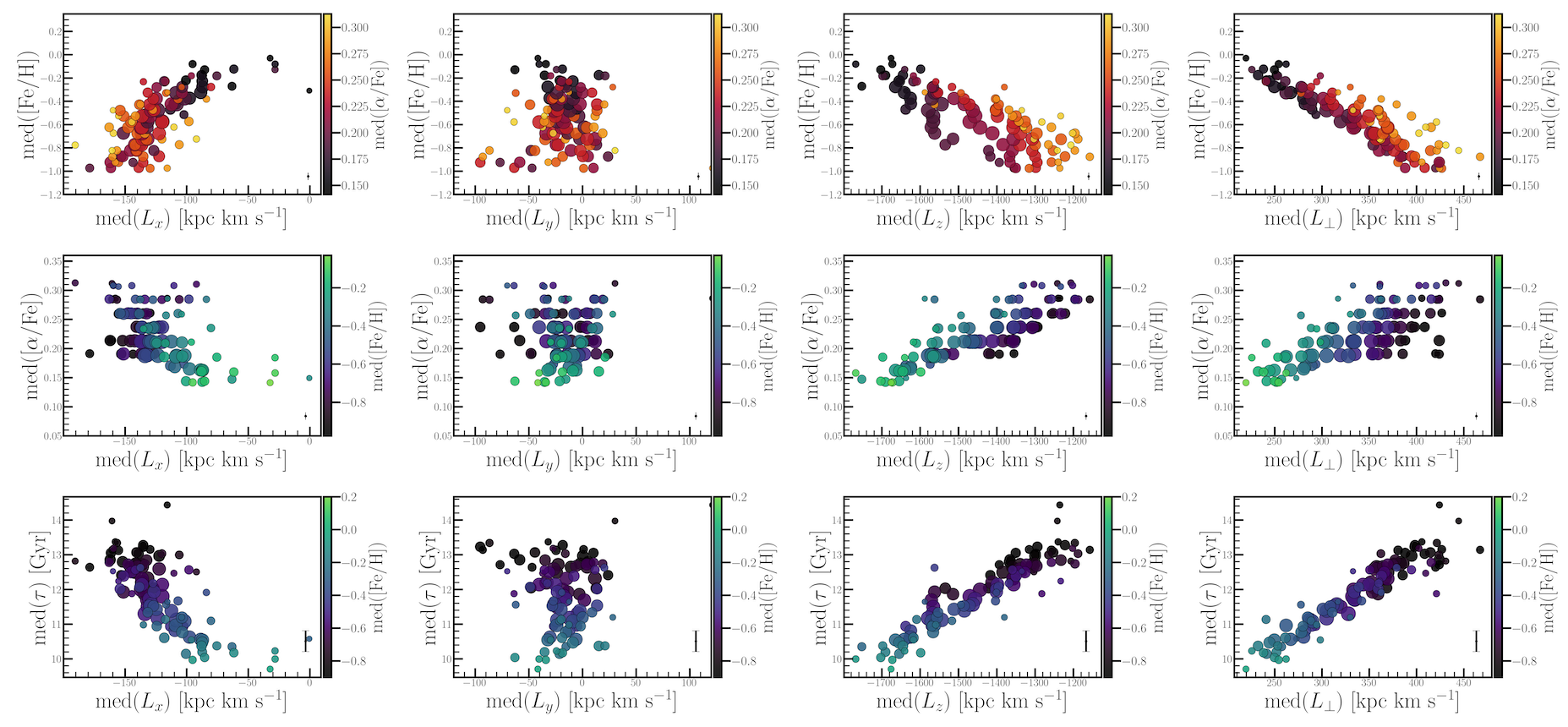}
    \caption{Correlations between chemical abundances, stellar ages, and angular momentum components for high-$\alpha$ stars. The top row shows the relation between [Fe/H] and the angular momentum components $L_x$, $L_y$, $L_z$, and $L_\perp$, while the middle row presents the corresponding relations for [$\alpha$/Fe]. The bottom row displays the dependence of stellar age $\tau$ on the same angular momentum components. Marker colors indicate the complementary abundance ([$\alpha$/Fe] in the top row, [Fe/H] in the middle row) and the [Fe/H] ratio in the bottom row, while marker sizes scale with the number of stars in each mono-abundance bin. Typical uncertainties of the median values, estimated as $\sigma/\sqrt{n}$ for each mono-abundance bin, are shown as representative error bars in the lower-right corner of each panel.}
    \label{fig:subgiants_ang_mom_map}
\end{figure*}

Following the examination of MAPs in action space, we now apply the same analysis with angular momentum. Figure~\ref{fig:subgiants_ang_mom_map} presents the variations of median [Fe/H], median [$\alpha$/Fe], and median stellar age as a function of the angular momentum components $L_x$, $L_y$, $L_z$, and their combined perpendicular component $L_\perp$ ($L_\perp = \sqrt{L_x^2 + L_y^2}$), using representative values computed within each MAP bin.

Overall we find that the trends of median element abundance and age are still present when examining the data as a function of $L_x,L_z,$ and $L_\perp$, but are non-existent with respect to $L_y$. In more detail, we find that: $i$) there is a positive trend between med([Fe/H]) and $L_x$, but a negative trend between med([$\alpha$/Fe]) and med(age) with $L_x$; $ii$) there is a strong positive(negative) trend between med([$\alpha$/Fe]) and med(age)(med([Fe/H])) and $L_z$, such that stars with smaller azimuthal angular momentum (in our convention, more positive values) tend to be on average more metal-poor, more [$\alpha$/Fe]-enhanced, and older; $iii$) there is a negative gradient in med([Fe/H]) with $L_\perp$, but a positive gradient in med([$\alpha$/Fe]) and med(age) with $L_\perp$. As observed in Fig~\ref{fig:subgiants_actions_map}, the median [$\alpha$/Fe] and median age follow similar profiles in angular momentum, that are inverted when examining med([Fe/H]). This inverted behavior suggests that the relative contributions of core-collapse supernovae (SN~II) and Type~Ia supernovae to the chemical enrichment of high-$\alpha$ disc stars vary with orbital structure.

Interestingly, we find that the distribution of MAPs in $L_x$ is not centered around $L_x=0$, as would be expected for a system in equilibrium (see also Fig~\ref{fig:subgiant_angularmomentum}). This could be due to effects in the selection function of \textsl{LAMOST} or our sample, or it could possibly be an indication that there is a coherent warp in the high-$\alpha$ disc. We are currently investigating the nature of this feature further with a much larger and complete sample of high-$\alpha$ disc stars (Akbaba et al, in prep). Lastly, as done for the previous section, we quantify the trends from Figure~\ref{fig:subgiants_ang_mom_map} in Table~\ref{tab:gradients}.

\begin{table*}[ht]
\centering
\renewcommand{\arraystretch}{1.3}
\caption{Linear regression gradients and bias terms for [Fe/H], [$\alpha$/Fe], and age ($\tau$)
         as a function of angular momenta ($L$) and orbital actions ($J$) in MAPs.
         Uncertainties represent $1\sigma$ errors from the regression. Gradient units are dex\,$(10^3\,\mathrm{kpc\,km\,s^{-1}})^{-1}$ for [Fe/H] and [$\alpha$/Fe], and $\mathrm{Gyr}\,(10^3\,\mathrm{kpc\,km\,s^{-1}})^{-1}$ for $\tau$. Bias terms are in dex for [Fe/H] and [$\alpha$/Fe], and in Gyr for $\tau$. Uncertainties are $1\sigma$ errors from ordinary least-squares regression. The perpendicular components are defined as the scalar sum of the in-plane ($x$--$y$ plane) contributions: $L_\perp = \sqrt{L_x^2 + L_y^2}$ and $J_\perp = \sqrt{J_R^2 + J_z^2}$.Gradient and bias uncertainties are derived analytically from ordinary least-squares (OLS) regression, following the standard covariance matrix formalism: $\sigma_{\hat{\beta}} = \sqrt{\hat{\sigma}^2 \left(\mathbf{X}^\top \mathbf{X}\right)^{-1}_{ij}}$, where $\hat{\sigma}^2$ is the residual variance and $\mathbf{X}$ is the design matrix.}
\label{tab:gradients}
\begin{tabular}{l|
                S[table-format=-2.3] @{$\,\pm\,$} S[table-format=1.3]
                S[table-format=-2.3] @{$\,\pm\,$} S[table-format=1.3]
                S[table-format=-1.3] @{$\,\pm\,$} S[table-format=1.3]
                S[table-format=-1.3] @{$\,\pm\,$} S[table-format=1.3]
                S[table-format=-1.3]  @{$\,\pm\,$} S[table-format=1.3]
                S[table-format=-1.3]  @{$\,\pm\,$} S[table-format=1.3]}
\toprule
 & \multicolumn{4}{c}{[Fe/H]}
 & \multicolumn{4}{c}{[$\alpha$/Fe]}
 & \multicolumn{4}{c}{$\tau$ (Gyr)} \\
\cmidrule(lr){2-5}\cmidrule(lr){6-9}\cmidrule(lr){10-13}
  & \multicolumn{2}{c}{Gradient}
  & \multicolumn{2}{c}{Bias}
  & \multicolumn{2}{c}{Gradient}
  & \multicolumn{2}{c}{Bias}
  & \multicolumn{2}{c}{Gradient}
  & \multicolumn{2}{c}{Bias} \\
\midrule
$L_x$
  &   6.772 & 0.626 &   0.289 & 0.079
  &  -0.682 & 0.080 &   0.122 & 0.010
  & -29.824 & 2.139 &   7.988 & 0.261 \\
$L_y$
  &   0.003 & 0.820 &  -0.558 & 0.024
  &   0.322 & 0.105 &   0.208 & 0.004
  &   0.536 & 3.913 &  11.572 & 0.103 \\
$L_z$
  &  -1.220 & 0.102 &  -2.339 & 0.149
  &   0.250 & 0.012 &   0.576 & 0.018
  &   6.256 & 0.214 &  20.928 & 0.322 \\
$L_\perp$
  &  -3.961 & 0.177 &   0.760 & 0.060
  &   0.614 & 0.048 &   0.003 & 0.016
  &  16.696 & 0.493 &   6.239 & 0.159 \\
\midrule
$J_R$
  &  -4.444 & 0.338 &  -0.091 & 0.036
  &   0.957 & 0.061 &   0.112 & 0.006
  &  23.562 & 0.881 &   9.358 & 0.085 \\
$J_z$
  & -12.249 & 0.690 &  -0.038 & 0.030
  &   2.298 & 0.156 &   0.114 & 0.006
  &  59.033 & 1.630 &   9.333 & 0.064 \\
$J_\phi$
  &   1.130 & 0.102 &  -2.155 & 0.146
  &  -0.256 & 0.012 &   0.577 & 0.018
  &  -6.096 & 0.224 &  20.454 & 0.330 \\
$J_\perp$
  &  -4.081 & 0.270 &  -0.030 & 0.035
  &   0.834 & 0.054 &   0.104 & 0.007
  &  20.755 & 0.585 &   9.159 & 0.070 \\[4pt]
\bottomrule
\end{tabular}
\end{table*}

\section{Discussion}
\label{sec:discussion}
\subsection{Element abundance and age orbital gradients in the Galactic high-$\alpha$ disc}

Our results (Fig~\ref{fig:kinematicsvsorbits} through Fig~\ref{fig:subgiants_ang_mom_map}) demonstrate that the median [Fe/H], [$\alpha$/Fe], and age of high-$\alpha$ disc stars exhibit a strong and ordered relation with their orbital structure. These trends are much clearer as a function of orbital properties than they are in kinematic diagrams (Fig~\ref{fig:kinematicsvsorbits}). In particular, the trends seen for med([Fe/H]), med([$\alpha$/Fe]), and med(age) across orbital actions (Figure~\ref{fig:subgiant_actions} and Fig~\ref{fig:subgiants_actions_map}) and angular momentum (Fig~\ref{fig:subgiant_angularmomentum} and Fig~\ref{fig:subgiants_ang_mom_map}) indicate that this population does not originate from a dynamically random or chaotic history. Instead, these trends reveal a tight coupling between chemical enrichment and orbital structure, pointing to a coherent chemical-orbital evolution of the high-$\alpha$ disc. Trends of element abundance and age with azimuthal and perpendicular values of angular momentum have been shown to exist in the high-$\alpha$ disc \citep{Hu2023}, again attesting to our results and the ordered structure of this Galactic component. 

Perhaps one of the most interesting results from this work has been the finding that, not only does the high-$\alpha$ disc display median [Fe/H], [$\alpha$/Fe], and age trends, but that these gradients are qualitatively similar to the low-$\alpha$ population (Fig~\ref{fig:kinematicsvsorbits_lowalpha} through Fig~\ref{fig:subgiant_angularmomentum_lowalpha}), albeit with weaker gradients. In the high-$\alpha$ disc, older, more [$\alpha$/Fe]-rich, and more [Fe/H]-poor populations have larger radial and vertical orbital excursions than their younger, more [$\alpha$/Fe]-poor, and more [Fe/H]-rich counterparts that follow more circular orbits (with higher $J_\phi/L_z$). This is the case when examining the data in full or when using MAPs. 

Our results are fully consistent with previous observational chemical-orbital studies in the literature \citep[e.g.,][]{Bovyetal2012,Rix2013, Minchev2013,Minchev2014,Mackereth2017, Frankel2019,Guctekin2019,Karaali2019,Horta2022,Imig2023}. One interesting comparison example is the work from \citet{Binney2024}, that used \textsl{APOGEE} DR17 data and analyzed the action distributions of high-$\alpha$ disc stars; their results illustrate a weak but systematic metallicity gradient along $J_\phi$, such that $J_\phi$ decreases with decreasing average [Fe/H]. They further showed that the vertical action $J_z$ increases as average [Fe/H] decreases, and that stars with [Fe/H] $< -0.8$~dex preferentially occupy regions of relatively high $J_z$ and a restricted range of $J_R$. The MAP-based results presented in this work are in strong agreement with these observational findings and other works \citep{Bovy2012b, Bovy2016,Minchev2017, Mackereth2019,Lian2022}.

One interesting interpretation of our results is the notion that, although the geometrically defined thick disc is commonly referred to as the chemically defined high-$\alpha$ disc \citep{Haywood2013, Haywood2015, Kordopatis_2015}, these two definitions cannot refer to the same stellar population \citep{Khoperskov2025}. While both definitions share similar properties, the geometrically defined thick disc hosts stars that are relatively young \citep{Martig2016, Mackereth2019,Lian2025}, which is not the case for the high-$\alpha$ disc (despite a small fraction of peculiar stars, see for example \citealp{Martig2015}). Moreover, scale-height measurements of [$\alpha$/Fe]-[Fe/H] MAPs in the Galactic disc show a smooth decreasing trend \citep{Bovyetal2012}, instead of a clear bi-modality as expected if two distinct geometrical discs were present. Given that element abundances and stellar ages are a more pristine tracer of the origin and nature stellar populations when compared to kinematic/orbital quantities that can vary over time due to changes in the Galactic potential, a definition of an old, less chemically evolved, and (on average) kinematically hotter population referred to as the high-$\alpha$ disc seems more logical than a geometrically defined component.

In summary, our results corroborate previous findings, and strongly support a picture in which the high-$\alpha$ disc acquired its chemical-orbital-age structure early in the Galaxy's history. Subsequent evolution appears to have largely preserved this ordered configuration; these results place important constraints on the formation timescales and physical mechanisms responsible for the origin of the Galactic thick disc, including its ``birth'' or ``spin-up'' phase \citep{Belokurov2022,Conroy2022, Chandra2024,Viswanathan2025}.

\subsection{The formation pathway of the high-$\alpha$ disc: upside-down and inside-out?}

The formation pathway of the high-$\alpha$ disc has long been a subject of debate \citep{Chiappini1997,Abadi2003, Brook2004, Kazantzidis2008, Villalobos2008, Bournaud2009, Schonrich2009, Loebman2011}. Some cosmological simulations and observational studies suggest that the high-$\alpha$ disc assembled both by cooling vertically and by growing radially outwards \citep{Bird2013, Minchev2014}. The element abundance and age gradients with action and angular momentum presented in Section~\ref{sec:results} provide new and direct observational constraints on this formation pathway. 

The strongest evidence for an upside-down formation comes from the vertical action, $J_z$, and the perpendicular angular momentum, $L_\perp$. Among all orbital components, $J_z$ displays the most ordered age gradient (Fig.~\ref{fig:subgiants_actions_map}, lower panels; $\mathrm{d}\tau/\mathrm{d}J_z = 59.0 \pm 1.6~\mathrm{Gyr}~(10^3~\mathrm{kpc~km~s^{-1}})^{-1}$), with MAP median ages rising continuously from $\sim$10~Gyr at $J_z \sim 20~\mathrm{kpc~km~s^{-1}}$ to $\sim$14~Gyr at $J_z \sim 100~\mathrm{kpc~km~s^{-1}}$. The median $L_\perp$--age relation ($\mathrm{d}\tau/\mathrm{d}L_\perp = 16.7 \pm 0.5~\mathrm{Gyr}~(10^3~\mathrm{kpc~km~s^{-1}})^{-1}$; Fig.~\ref{fig:subgiants_ang_mom_map}, lower panels) independently corroborates this finding. \citet{Bird2013} analysed mono-age cohorts in the Eris cosmological simulation and showed that the kinematic properties of each cohort are largely imprinted at birth or immediately thereafter: the oldest cohorts form in a kinematically hot and vertically extended environment, while subsequent cohorts originate from progressively thinner and cooler configurations. The smooth and uninterrupted $J_z$--age relation we recover is in direct agreement with this picture. Such continuity is also inconsistent with a scenario in which the vertical structure of the high-$\alpha$ disc was established through a single impulsive merger-induced heating event, which would instead be expected to leave a sharp discontinuity at a characteristic age in the diagram. The $J_z$--age relation recovered in our high-$\alpha$ sample shows no evidence for such a discontinuity. Median stellar ages increase smoothly and continuously from $\sim$10 Gyr to $\sim$14 Gyr across the full range of vertical action (Fig.~\ref{fig:subgiants_actions_map}), without any detectable gap or change in slope at a characteristic age. This continuity argues against a scenario in which the vertical structure of the high-$\alpha$ disc was established through temporally separated formation episodes.

To place our results in a broader context, it is instructive to compare them with recent observational evidence for multiple thick-disc components. \citet{Lian2025} recently identified two geometrically thick disc populations in the Milky Way using \textsl{APOGEE} and \textsl{Gaia} data: a canonical old thick disc ($\sim$9.3~ Gyr, $h_z \sim 0.77$ kpc), associated with the high-$\alpha$ sequence, and a younger geometrically thick component ($\sim$6.6 Gyr) with lower $\alpha$ enhancement. These two populations are separated by a clear density gap at [Mg/Fe] $\approx 0.2$~dex, suggesting that the Milky Way experienced at least two discrete episodes of turbulent, upside-down disc formation. A natural interpretation is that the two-component structure reported by \citet{Lian2025} reflects that a geometrical selection of the Milky Way disc directly leads to a sample comprised of high- and low-$\alpha$ stars; thus, the evidence for a separate geometrically defined thick disc that isolates the high-$\alpha$ population is not well founded. Instead, a selection of a high-$\alpha$ population, as in our analysis, leads to a population with a structure that appears to be consistent with a single, extended phase of upside-down and inside-out formation. 


Independent and complementary evidence for inside-out growth comes from the $J_\phi$--age and $L_z$ age relations. Stellar ages decrease monotonically from $\sim$13--14~Gyr at $J_\phi \sim 1100-\mathrm{kpc~km~s^{-1}}$ to $\sim$10~Gyr at $J_\phi \sim 1700~ \mathrm{kpc~km~s^{-1}}$ (Fig.~\ref{fig:subgiants_actions_map}, lower panels), with the same trend observed in $L_z$--age space (Fig.~\ref{fig:subgiants_ang_mom_map}, lower panels). 

\citet{Martig2016} analysed $\sim$14,685 red giant stars from \textsl{APOGEE} DR12 and showed that the geometrically defined thick disc exhibits a pronounced radial age gradient, whereas the chemically defined ($\alpha$-rich) thick disc displays an age distribution that is largely independent of Galactocentric radius. This discrepancy may be a direct consequence of examining older (more dynamically heated) populations in kinematic space, which we have seen is less informative than in orbital diagrams (Fig~\ref{fig:kinematicsvsorbits}). The $J_\phi$--age gradient recovered in action space advocates for the orbital analysis to this discussion, providing a direct imprint of inside-out growth within the high-$\alpha$ disc that is not captured in kinematic space. Along these lines, \citet{Hu2023} derived a similar inside-out growth timescale of $\mathrm{d}R_g/\mathrm{d}\tau = -1.9~\mathrm{kpc~Gyr^{-1}}$ for the high-$\alpha$ disc as the one found in this study using \textsl{APOGEE} data, corroborating our results.

The $J_\phi$--$[\mathrm{Fe/H}]$ relation provides additional constraints on the structured evolution of the high-$\alpha$ disc. The positive gradient we recover ($\mathrm{d}[\mathrm{Fe/H}]/\mathrm{d}J_\phi = +1.13 \pm 0.10~\mathrm{dex}~(10^3~\mathrm{kpc~km~s^{-1}})^{-1}$) --- reflecting the tendency of stars on more circular orbits to be more metal-rich --- is a natural outcome of inside-out and upside-down formation. \citet{Kawata2018} combined N-body simulations with \textsl{APOGEE} data and showed that the initial radial metallicity gradient of the thick disc progenitor must have been flat or positive, and that the combination of inside-out growth and an age--metallicity relation naturally produces the positive $v_\phi$--$[\mathrm{Fe/H}]$ slope observed for thick disc stars. Our $J_\phi$--$[\mathrm{Fe/H}]$ gradient is in direct agreement with this prediction.

Finally, it is worth noting that the regularity of the age--action correlations is not uniform across all orbital components. While $J_R$, $J_z$, $J_\phi$, $J_\perp$, $L_z$ and $L_\perp$ all display tight and well-ordered relations with age, the in-plane components $L_x$ and $L_y$ show considerably more scatter despite statistically significant gradients. This is expected on dynamical grounds: $J_R$, $J_z$ and $J_\phi$ are adiabatic invariants approximately conserved in an axisymmetric potential, and therefore retain a memory of the kinematic conditions at the time of a star's formation. In contrast, $L_x$ and $L_y$ are only conserved under full spherical symmetry, progressively washing out any formation signal. The fact that this excess scatter does not propagate into the $J_z$--age and $J_R$--age relations suggests that the ordered chemical-orbital structure of the high-$\alpha$ disc has been largely preserved over long dynamical timescales. 

\subsection{Our results in the context of the Milky Way's accretion history}

Results from large-scale stellar surveys have revealed that the Milky Way has undergone a number of accretion events during its lifetime \citep[e.g.,][]{Koppelman2019,Myeong2019,Horta2021, Dodd2023, Horta2023}, with the most significant merger event experienced by the Milky Way postulated to be approximately $\approx8-11$ Gyr ago \citep[\textsl{Gaia-Enceladus/Sausage}, or GES:][]{Helmi2018, Belokurov2018, Gallart2019}. Of particular interest to the results from this study is this GES merger; if this accretion event was as significant as postulated in these discovery studies ---leading the formation of the inner halo and thick disc--- then it should have erased any ordered chemical--age--orbital structure in the oldest stars in the Milky Way disc. Such a major-merger scenario would naturally lead to element abundance and age gradients with respect to orbits to have been completely washed--out, or at the bare minimum, show a clear discontinuity in chemical-orbital and age-orbital trends for old high-$\alpha$ disc stars. However, the findings presented in this study advocate for a different scenario. Our results show that the old ($\tau >8$ Gyr) high-$\alpha$ disc is highly structured, where the persistence and clarity of the observed element abundance and age correlations suggest that the main body of the high-$\alpha$ disc survived merger events without being fully altered. Naturally, our findings would then imply that the GES merger, or any other accretion event afterwards, must not have been a significant merger with a large host-to-merger ratio.

This interpretation is consistent with recent cosmological simulation results \citep{Taylor2017}, especially for early-formed systems like the high-$\alpha$ disc \citep{Yoon2023}. For instance, \citet{Renaud2025} showed using the \texttt{VINTERGATAN} simulation that [$\alpha$/Fe] ratios may respond to short-lived starburst episodes potentially caused by merger events, even if signatures in [Fe/H] are not rapidly erased. The presence of clear gradients in [$\alpha$/Fe] implies that merger events were not significant enough to fully erase any $\alpha$-element and orbital gradient. Thus, the smooth variation of [$\alpha$/Fe] with orbital actions found in this work suggests that the chemical structure of the high-$\alpha$ disc was established early and subsequently preserved over long dynamical timescales. While the overall radial--metallicity gradient of populations formed after the merger can be sustained, any chemical-orbital or age-orbital gradients formed before would have been erased.


Along those lines, the GES merger has been conjectured to be the culprit of high-$\alpha$ disc stars on highly radial orbits (called the \textit{Splash}: \citealp{Bonaca2017, Belokurov2020}). While it is totally viable for the GES merger (or other accretion events) to have heated some high-$\alpha$ disc stars (\citealp{Kisku2025}, although see also \citet{Buder2025} and references therein), the fraction of stars comprising this population must be small compared to the overall high-$\alpha$ disc population. In this study, we have been unable to statistically identify these radially heated disc populations, and instead we just find a tail of more radial orbits in the high-$\alpha$ disc.

\section{Conclusions}
\label{sec:conclusions}

In this work, we have used the sample of subgiant stars with chemical, orbital, and age information from \citet{Xiang2022} to perform a chemical and age orbital cartography of the Galactic high-$\alpha$ disc. Our main findings can be summarised as follows: 

\begin{itemize}
    \item The median [Fe/H], median [$\alpha$/Fe], and median age of high-$\alpha$ disc stars display ordered trends with kinematics and orbits (Fig~\ref{fig:subgiant_actions} through to Fig.~\ref{fig:subgiants_ang_mom_map}). However, these trends are notably stronger in orbit space when compared to kinematic diagrams (Fig.~\ref{fig:kinematicsvsorbits}). The presence of element abundance and age trends with orbital actions or angular momenta reflects a tight coupling between chemical enrichment and orbital evolution, indicating that this population has an ordered structure that was set in the early Universe and has been preserved over time. 
    
    \item The element abundance- and age-orbit trends are qualitatively similar to those seen in the low-$\alpha$ disc (Fig.~\ref{fig:kinematicsvsorbits_lowalpha}--Fig~\ref{fig:subgiant_angularmomentum_lowalpha}), albeit with weaker gradients.

    \item The element abundance-orbit and age-orbit gradients observed in the high-$\alpha$ disc are consistent with an upside-down and inside-out formation scenario (see Table~\ref{tab:gradients} for a summary). 

    \item The persistence and clarity of the element abundance and age correlations with respect to orbital properties in the high-$\alpha$ disc suggest that the structure of the main body of this population survived the effects of the Milky Way's accretion history --- including the \textit{Gaia-Enceladus/Sausage} merger --- without being fully disrupted. This implies that this merger event, or any subsequent one, must not have had a large host-to-merger mass ratio, since such significant mergers would have completely erased the ordered chemical-age-dynamical structure of the old disc. The GES merger may have heated a fraction of high-$\alpha$ disc stars onto more radial orbits, producing the so-called \textit{Splash} population, but this fraction appears to be small relative to the overall high-$\alpha$ disc (Akbaba et al, in prep.).
\end{itemize}

Taken together, our results place new quantitative constraints on the formation timescales and physical mechanisms responsible for the origin of the Galactic high-$\alpha$ disc, and are broadly consistent with the predictions of simulations \citep{Bird2013, Minchev2014}. The ordered chemical-age-orbital structure of the high-$\alpha$ disc seen in this study underscores the power of combining stellar ages and element abundance information with Galactic dynamics to constrain the formation and evolution of some of the Galaxy's oldest stellar populations.

Looking ahead, the results presented in this work motivate several avenues for the future investigation. The origin of the offset in the $L_x$ distribution of high-$\alpha$ disc MAPs from zero (Fig~\ref{fig:subgiant_angularmomentum} --- which may reflect either a selection effect in the LAMOST sample or a coherent warp in the high-$\alpha$ disc --- will be explored further using a larger and more complete sample of high-$\alpha$ disc stars (Akbaba et al, in prep.). Furthermore, in this study we have solely focused on examining the two most well understood element abundances, namely [Fe/H] and [$\alpha$/Fe], and in a relatively small volume around the Sun. However, new surveys like \textsl{Milky Way Mapper/SDSS-V} \citep{Kollmeier2026} have and will collect high-resolution spectra for millions of stars across large swaths of the Milky Way disc; we aim to revisit this analysis with such data to analyse element abundance trends (using a spectrum of elemental species) across larger regions of the high-$\alpha$ disc.

\section*{Acknowledgements}

The authors would like to thank Alexander Stone-Martinez, Julie Imig, and Jorge Pen\~arrubia for helpful discussions.
This work was supported by the Scientific and Technological Research Council of Turkey (TÜBİTAK) under project code MFAG-123F227 and program 2211-C. This study was funded by the Scientific Research Projects Coordination Unit of the Istanbul University. Project numbers 40044, FBA-2023-39380, and FDK-2025-41537.
DH is supported by the UKRI Science and Technology Facilities Council under project 101148371 as a Marie Skłodowska-Curie Research Fellowship. 
\section*{Data Availability}

\textit{Data}: The LAMOST spectroscopic data used in this work are publicly available through the LAMOST data release portal at \url{http://www.lamost.org}. The \textit{Gaia} astrometric data are publicly available through the \textit{Gaia} Archive at \url{https://gea.esac.esa.int/archive}. The derived stellar ages used in this work are taken from \citet{Xiang2022}.\\
\textit{Software}: \texttt{astropy} \citep{Astropy2013, Astropy2018, astropy2022},  \texttt{NumPy} \citep{Numpy, Numpy2020}, \texttt{scikit-learn} \citep{scikit-learn}, \texttt{matplotlib} \citep{Matplotlib, Matplotlib2007}, \texttt{galpy} \citep{galpy}.



\bibliographystyle{mnras}
\bibliography{references} 



\appendix

\section{Scatter in element abundances and age across orbital diagrams}

\begin{figure*}
    \centering
    \includegraphics[width=\textwidth]{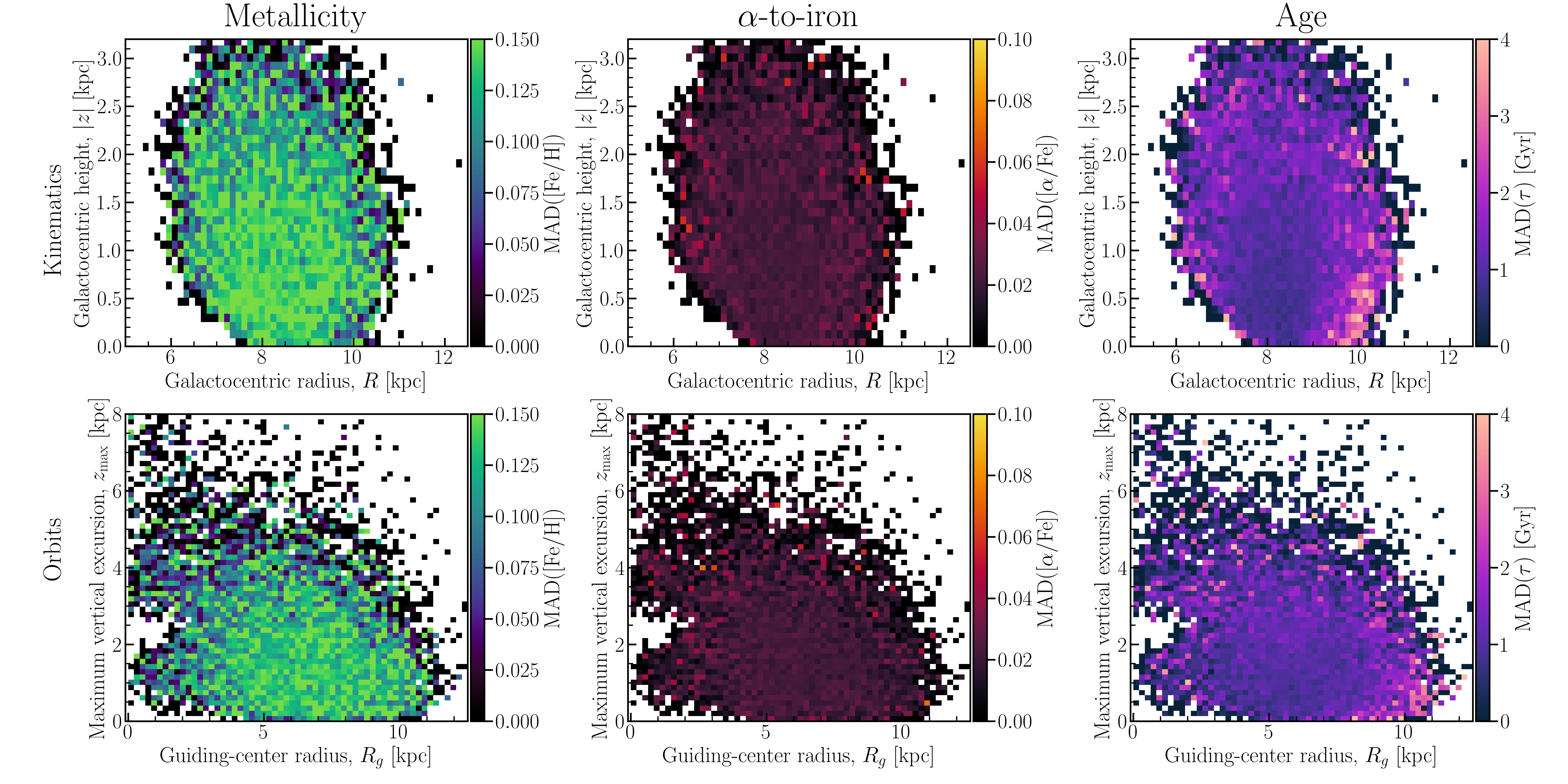}
    \caption{Spatial and orbital distributions of the high-$\alpha$ population. The top row shows maps in the $R$-$|z|$ plane and the bottom row shows the corresponding distributions in the $R_g$-$z_{\max}$ plane. Colours represent the median absolute deviation (MAD) of [Fe/H], [$\alpha$/Fe], and stellar age in each spatial bin, providing a robust estimate of the internal dispersion of these quantities.}
    \label{fig:kinematics_vs_orbits_MAD}
\end{figure*}

We have verified that the element abundance-orbital and age-orbital gradients reported in Section~\ref{sec:results} are not a product of internal scatter within each bin. Figure~\ref{fig:kinematics_vs_orbits_MAD} show the MAD of [Fe/H], [$\alpha$/Fe], and stellar age across the $R$--$|z|$ and $R_g$--$z_{\rm max}$ planes. The scatter is approximately uniform across all pixels, with typical values of $\approx0.13$ dex for [Fe/H], $\approx0.02$ dex for [$\alpha$/Fe], and $\approx1$ Gyr for age, confirming that the observed trends reflect genuine structural properties of the high-$\alpha$ disc rather than binning artefacts. 

\begin{figure*}
    \centering
    \includegraphics[width=\textwidth]{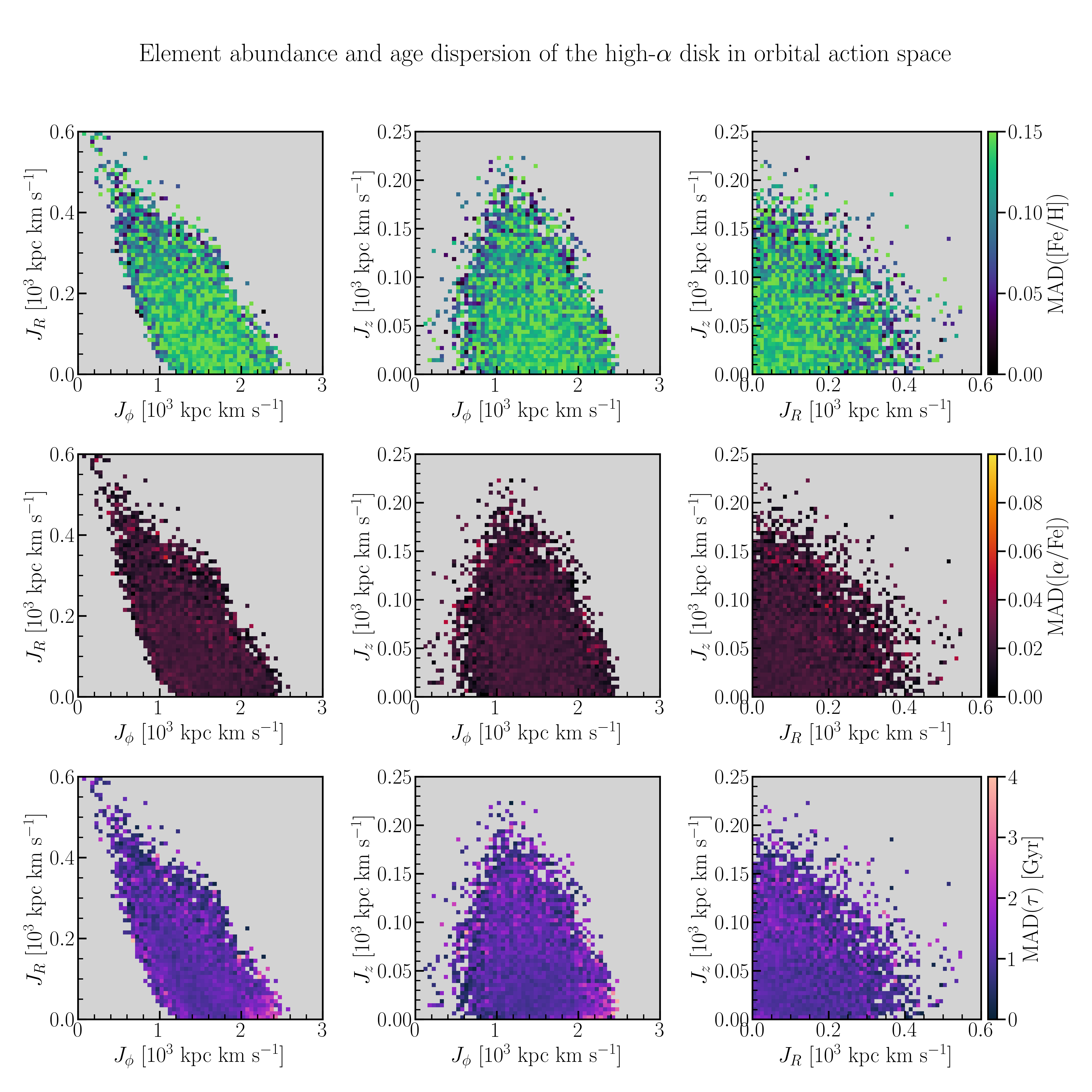}
    \caption{Distribution of high-$\alpha$ stars in action space. The panels show maps of the median absolute deviation (MAD) of [Fe/H] (top row), [$\alpha$/Fe] (middle row), and stellar age (bottom row) across different projections of the actions $(J_R, J_z, J_\phi)$. The MAD provides a robust estimate of the internal dispersion of these quantities within each bin, revealing how the chemical and age spreads vary across action space.}
    \label{fig:subgiants_actions_MAD}
\end{figure*}

\begin{figure*}
    \centering
    \includegraphics[width=\textwidth]{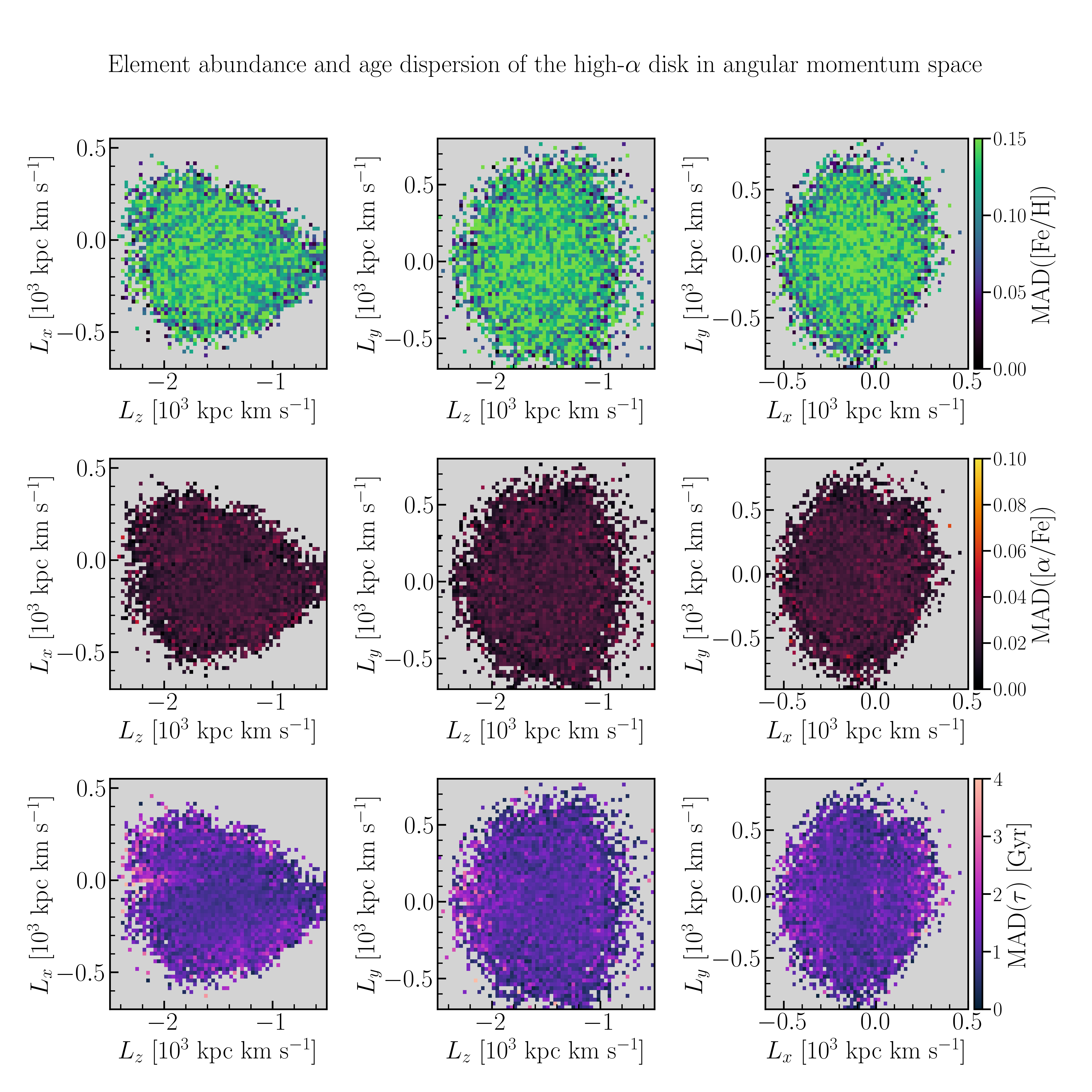}
    \caption{Distribution of high-$\alpha$ stars in angular momentum space. The panels show the median absolute deviation (MAD) of [Fe/H] (top row), [$\alpha$/Fe] (middle row), and stellar age (bottom row) across different projections of the angular momentum components $(L_x, L_y, L_z)$. Colors indicate the MAD of each quantity within spatial bins, providing a robust measure of the intrinsic dispersion and highlighting variations in the element abundance-orbital and age-orbital structure of the high-$\alpha$ population.}
    \label{fig:subgiants_angular_momentums_MAD}
\end{figure*}

Similarly, the internal scatter within each mono-orbit population in action space (Fig.~\ref{fig:subgiants_actions_MAD}) and angular momentum space (Fig.~\ref{fig:subgiants_angular_momentums_MAD}) is small and varies smoothly across all orbital components. These results confirm that the high-$\alpha$ disc is a structured population, and that the median trends in element abundances and stellar age as a function of orbital properties are robust.

\section{chemical-orbital trends of the low-$\alpha$ disc in action space}
\label{app_lowalpha}
\begin{figure*}
    \centering
    \includegraphics[width=\textwidth]{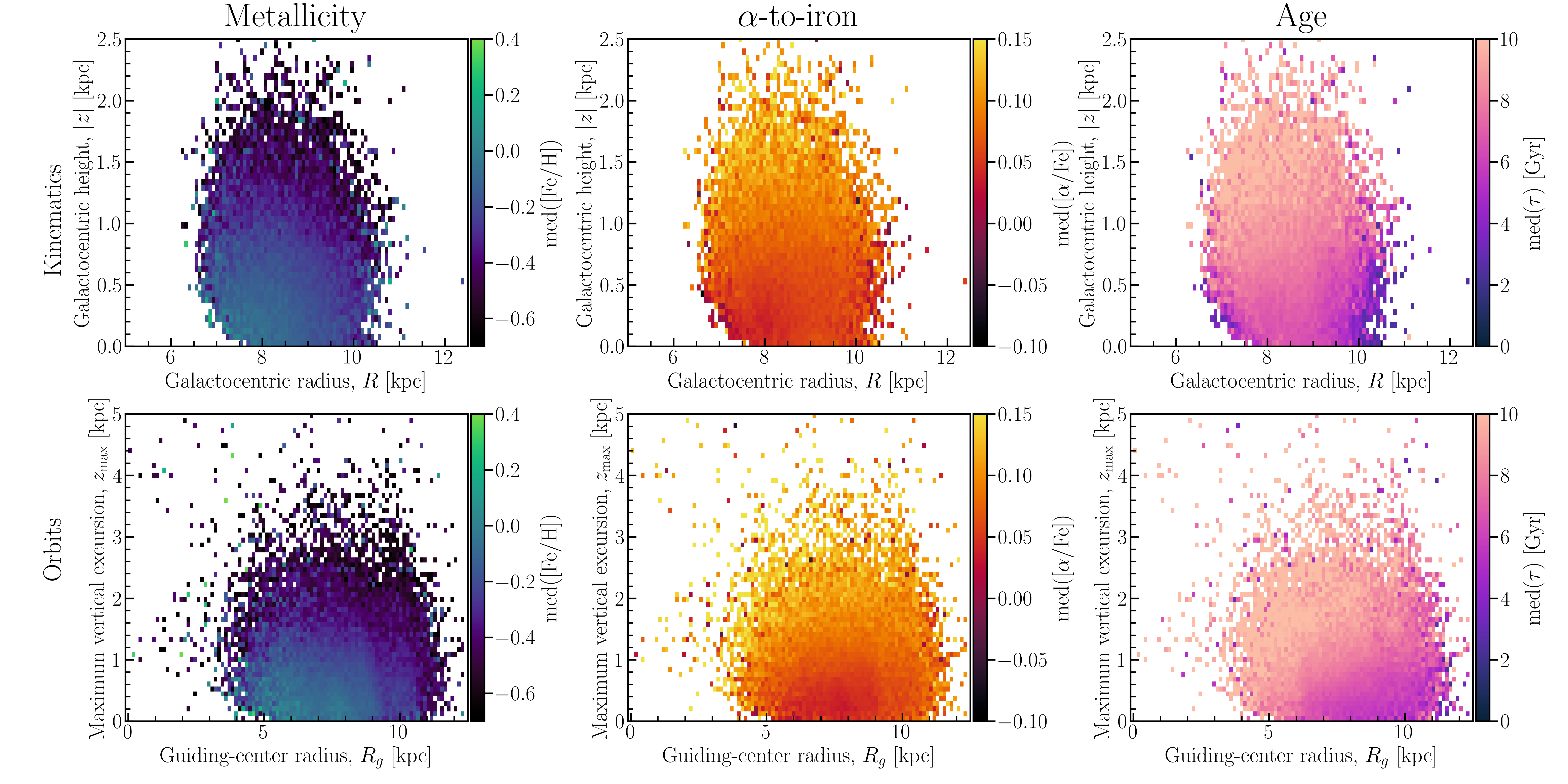}
    \caption{Kinematic and orbital distributions of the low-$\alpha$ population. The top row shows maps in the $R$-$|z|$ plane, while the bottom row displays the corresponding distributions in the $R_g$-$z_{\max}$ plane. From left to right, panels are colour-coded by median of [Fe/H], [$\alpha$/Fe], and stellar age.}
    \label{fig:kinematicsvsorbits_lowalpha}
\end{figure*}

\begin{figure*}
    \centering
    \includegraphics[width=\textwidth]{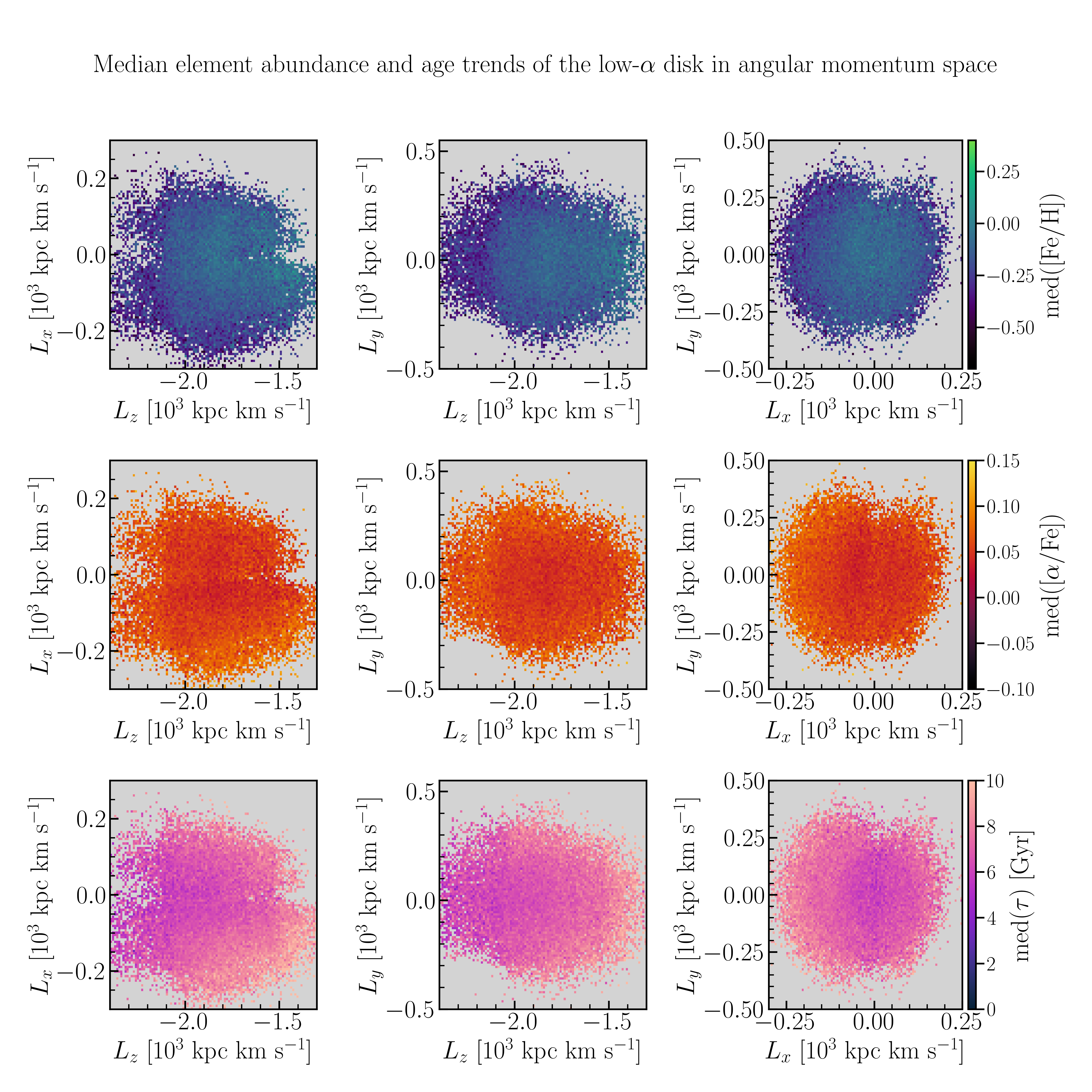}
    \caption{Distribution of low-$\alpha$ stars in angular momentum space. The panels show the median values of [Fe/H] (top row), [$\alpha,$/Fe] (middle row), and stellar age (bottom row) across different projections of the angular momentum components $(L_x, L_y, L_z)$. Colors indicate the median chemical abundance or median age within each bin, highlighting the element abundance--orbital and age--orbital structure of the low-$\alpha$ disc population.}
\label{fig:subgiant_angularmomentum_lowalpha}
\end{figure*}

For comparison with the high-$\alpha$ disc results  presented in the main body of this work, Figure~\ref{fig:kinematicsvsorbits_lowalpha} throught to Fig~\ref{fig:subgiant_angularmomentum_lowalpha} shows the distribution of low-$\alpha$ disc stars in kinematic, orbital action, and angular momentum space, colour-coded by median of [Fe/H], [$\alpha$/Fe], and stellar age. 

Qualitatively, similar element abundance-orbital and age-orbital trends are present in both populations; as expected, these trends appear more clearly defined in the low-$\alpha$ disc. The median stellar ages of the low-$\alpha$ population are also systematically younger than those of their high-$\alpha$ counterparts.

This comparison underscores the more tightly coupled and ordered element abundance-orbital and age-orbital structure of the high-$\alpha$ disc, and is consistent with our conclusion that this structure was shaped by a more rapid early formation history relative to the more extended assembly of the low-$\alpha$ population.

\begin{figure*}
    \centering
    \includegraphics[width=\textwidth]{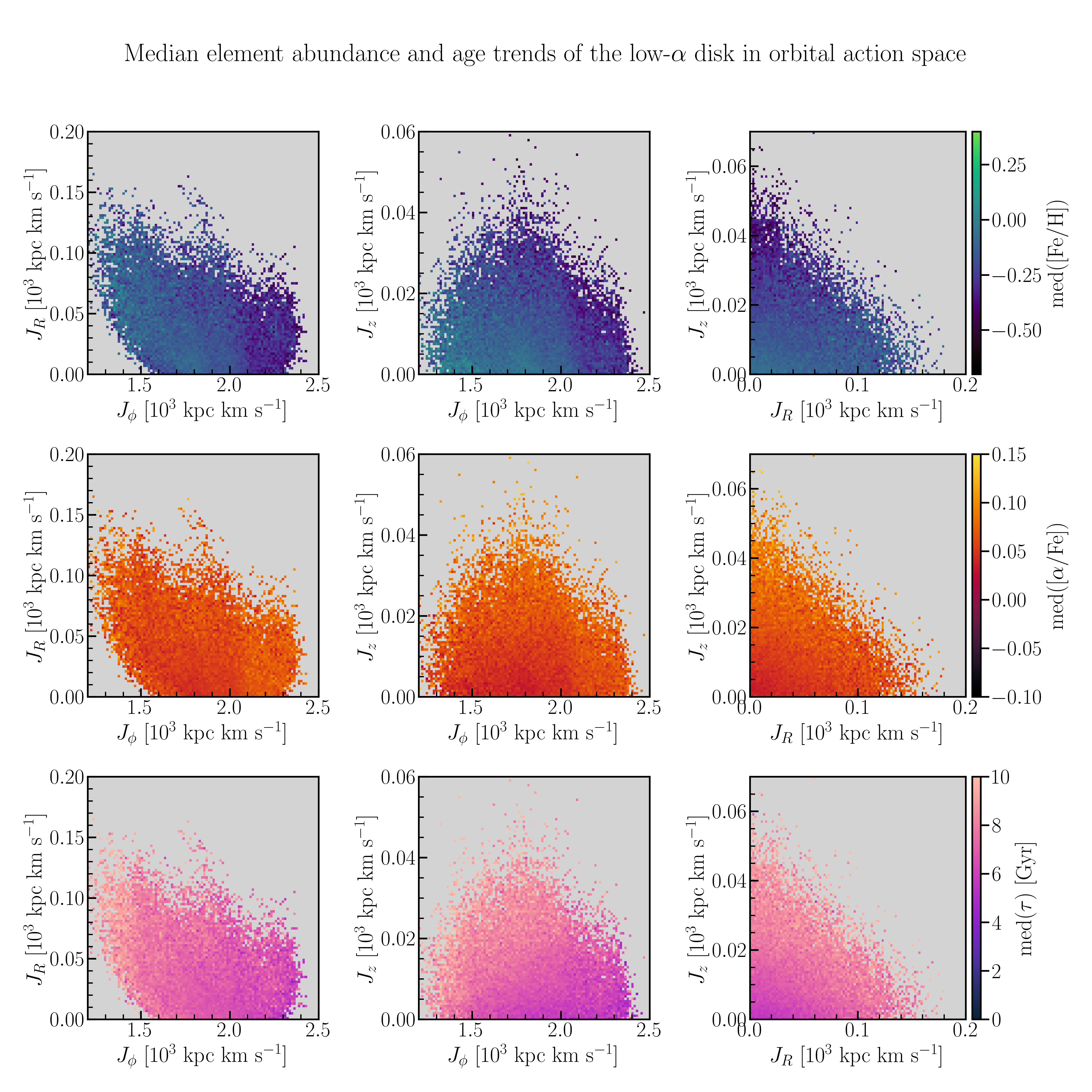}
    \caption{Distribution of low-$\alpha$ stars in action space. The panels show the median values of [Fe/H], [$\alpha$/Fe], and stellar age across different projections of the action plane. Colors indicate the average chemical abundance or age within each bin, highlighting systematic trends in the element abundance-orbital structure of the high-$\alpha$ population.}
    \label{fig:lowalpha_action}
\end{figure*}

\bsp	
\label{lastpage}
\end{document}